# Physical Characterization of Near-Earth Asteroid (52768) 1998 OR2: Evidence of Shock Darkening/Impact Melt

Characterization of NEA (52768) 1998 OR2

Adam Battle[1], Vishnu Reddy[1], Juan A. Sanchez[2], Benjamin Sharkey[1], Neil Pearson[2], Bryn Bowen[1]

[1] Lunar and Planetary Laboratory, University of Arizona, 1629 E. University Boulevard, Tucson, AZ 85721-0092, USA

[2] Planetary Science Institute, 1700 East Fort Lowell Road, Tucson, AZ 85719, USA

Abstract:

We conducted photometric and spectroscopic characterization of near-Earth asteroid (52768) 1998 OR2 during a close approach to the Earth in April of 2020. Our photometric measurements confirm the rotation period of the asteroid to be 4.126 ± 0.179 hours, consistent with the previously published value of 4.112 ± 0.001 hours. By combining our visible spectroscopic measurements (0.45–0.93 µm) with archival MITHNEOS near infrared spectra (0.78–2.49 µm), we classify the asteroid as an Xn-type in the Bus-DeMeo taxonomy. The combined spectrum shows two weak absorption bands: Band I at 0.926 ± 0.003 µm and Band II at 2.07 ± 0.02 µm with band depths of 4.5 ± 0.15% and 4.0 ± 0.21%, respectively. The band area ratio is 1.13 ± 0.05. These spectral band parameters plot at the tip of the S(IV) region of the Gaffey S-asteroid subtypes plot suggesting an affinity to ordinary chondrite meteorites. We calculated the chemistry of the olivine and pyroxene using the Band I center to be 20.1 ± 2.3 mol% fayalite and 18.2 ± 1.5 mol% ferrosilite, consistent with H chondrites. Principal component analysis of 1998 OR2's combined visible-NIR spectrum fall on the C/X-complex side of the α-line, near the end of the shock darkening trend, consistent with its weak absorption bands (band depth < 5%). We use an aerial mixing model with lab measurements of the shock darkened H5 chondrite, Chergach, to constrain the amount of shock darkened material on the asteroid's surface at ~63% dark lithology and ~37% light lithology.

1. INTRODUCTION

S-complex asteroids dominate the km-sized near-Earth asteroid (NEA) population and make up the majority of NEAs at all sizes (DeMeo et al. 2009; Binzel et al. 2019; Ieva et al. 2020). Several studies have linked these asteroids to ordinary chondrite meteorites (e.g., Gaffey et al. 1993; Vernazza et al. 2008; Binzel et al. 2009, 2019; DeMeo et al. 2009; Nakamura et al. 2011; Thomas et al. 2011). Spectral band parameters measured from visible and near infrared (VNIR; 0.45 – 2.5 µm) spectra of S-type asteroids are diagnostic of the surface mineralogy of these asteroids (Reddy et al. 2015). This allows more in-depth and quantitative comparisons of asteroid surfaces and their proposed meteorite analogs. However, non-compositional effects including grain size, phase angle, and alteration mechanisms such as space weathering and impact processes, can affect our ability to make these asteroid-meteorite linkages.

Lunar-style space weathering is an alteration of the surface of airless bodies that occurs due to long periods of time exposed to the space environment (Gaffey 2010). Nanophase iron particles created from micrometeorite impacts, solar wind implantation, and other space environment factors are the main source of the effects of lunar-style space weathering as described in Pieters & Noble (2016) and references within. These effects have been studied for over 50 years, since before the Apollo Moon landing (Zeller & Ronca 1967), but shock darkening and impact melt alteration processes have not been studied as long, with one of the first references being Britt & Pieters (1989). Whereas studies of space weathering persisted throughout the development of asteroid science (Hapke 2001; Brunetto et al. 2006; Hiroi et al. 2006; Gaffey 2010; Noguchi et al. 2011), studies of shock darkening and impact melt only picked up again after the 2013 Chelyabinsk event (Kohout et al. 2014, 2020a, 2020b; Reddy et al. 2014).

Space weathering affects reflectance spectra of S-type asteroids primarily by lowering the visual albedo, suppressing absorption bands, and reddening their VNIR spectrum (Hiroi et al. 2006; Gaffey 2010; Noguchi et al. 2011; Pieters & Noble 2016). Shock darkening also reduces the albedo and suppresses mineralogical band features, but typically does not redden the spectrum (Kohout et al. 2014). Shock darkening also changes physical aspects of meteorites and asteroid regolith, such as porosity and magnetic susceptibility, which cannot be detected through current remote sensing techniques (Kohout et al. 2014, 2020b, 2020a; Reddy et al. 2014).

Shock darkening observed in the meteorite collection is the result of partial or complete troilite or FeNi metal melting and mobilization due to impact-related heating. Two stages of shock darkening are observed at different pressure ranges. The first stage with lower amounts of darkening occurs at ~40-60 GPa with injection of melts into silicate grains. The second stage occurs at pressures between 90-150 GPa where bulk melting takes place and troilite and metal are finely dispersed within the silicate material. The latter is sometimes referred to as impact melt to differentiate the two phases of shock darkening (Kohout et al. 2020a, 2020b). The two phases cannot effectively be distinguished with remote sensing and we primarily use the term shock darkening to refer to both types of impact heating-related darkening of an asteroid surface.

Analysis of shock darkening in meteorites has been used to derive trends in principal component space that can help distinguish between space weathering and shock darkening (e.g., Binzel et al.

2019). Where space weathering is believed to alter asteroid spectra from Q-type to S-type, shock darkening and impact melt trends observed in meteorite spectra would result in asteroid spectra moving from S- or Q-type to the C/X-complex in principal component space (DeMeo et al. 2009; Kohout et al. 2014; Reddy et al. 2014; Binzel et al. 2019).

Here we show spectroscopic measurements of asteroid (52768) 1998 OR2 as evidence of possible shock darkened material on the surface of an asteroid in the NEA population. This implies that some S-complex NEAs may be altered by shock processes into resembling spectral properties of lower-albedo NEAs.

1.1 Background on 1998 OR2

1998 OR2 is an Apollo class asteroid classified as a potentially hazardous asteroid [JPL Small Body Database]. The most recent results from the NEOWISE survey of 1998 OR2 use an absolute magnitude, H = 16.0, and estimate a diameter of 2.51 ± 0.81 km and a visible geometric albedo of $0.164^{+0.123}_{-0.07}$ (Masiero et al. 2021). Arecibo radar measurements give a diameter of approximately 2.16 km (Devogèle et al. 2020). The radar estimated diameter can be used in the diameter-albedo relationship from Pravec & Harris 2007 (see their equation 2) to confirm the albedo.

$$p_v = \left[\frac{1329}{D(km)}\right]^2 * 10^{-2H/5} \quad (1)$$

Using the Arecibo estimated 2.16 km diameter, the derived visible geometric albedo is 0.15 which is consistent with the NEOWISE values. These albedo values are lower than the average for S-types in the NEA population, $0.26^{+0.04}_{-0.03}$, and are closer to the low/moderate-albedo C-type average albedo of $0.13^{+0.06}_{-0.05}$ (Thomas et al. 2011). The albedo of 1998 OR2 is, however, consistent with the average main belt S-type albedo of 0.174 ± 0.039 found by Ryan & Woodward (2010).

The earliest rotational period estimate for 1998 OR2 was 3.198 ± 0.006 h from Betzler & Novaes (2009). Koehn et al. (2014) provided a new period estimate of 4.112 ± 0.002 h which has been confirmed with optical observations by Warner & Stephens (2020) [4.1114 ± 0.0002 h] and Franco et al. (2020) [4.111 ± 0.001 h] as well as sequences of delay-doppler images from Arecibo Observatory[1] [reported as 4.1 h] during the asteroid's close approach in 2020. Variations in lightcurve amplitudes are seen on different nights of 1998 OR2's close approach in Warner & Stephens (2020), likely due to rapidly changing viewing conditions of the topography on the asteroid. Most recently, Colazo et al. (2021) claims a 4.01 ± 0.02-h rotation period which does not agree with most previously published values.

2. PHOTOMETRIC STUDY

   2.1. Observations and Data Reduction

Photometric observations of 1998 OR2 were obtained at LEO Observatory (MPC Code V17) on 2020 April 16 and 17 UTC. These observations were made prior to the asteroid's close approach to Earth on 2020 April 29 UTC at a nominal distance of 0.042 AU or approximately 16 lunar distances [JPL Small Body Database]. The 0.52-m, f/2.9 telescope is equipped with a 4k x 4k, 9 µm-pixel detector and a filter wheel with Sloan g', r', i', and z' filters resulting in a field of view of 1.4 x 1.4 degrees with a 1.2" px$^{-1}$ plate scale. Images were taken with 30 s exposure times in each of the four Sloan filters. The data were calibrated using standard CCD reduction techniques of dark- and bias-subtraction with flat-field division (Howell 2006). The number of images in each filter and other observing conditions are shown in Table 1.

---
[1] http://www.naic.edu/~pradar/press/1998OR2.php

*Table 1*. Observational circumstances for photometric observations

| Telescope | Date UTC | Start UTC | End UTC | Airmass | V. Mag | α (deg) | Filters | No. of Img. (g', r', i', z') |
|---|---|---|---|---|---|---|---|---|
| LEO | 2020 April 16 | 02:44 | 07:03 | 1.01 – 2.06 | 12.98 | 105.6 | SDSS g', r', i', z' | 78, 77, 77, 77 |
| LEO | 2020 April 17 | 02:50 | 06:17 | 1.02 – 1.62 | 12.85 | 105.8 | SDSS g', r', i', z' | 74, 73, 74, 64 |

Photometric extraction was performed with a custom in-frame photometry pipeline written in Python which utilizes astropy, photutil, TKinter, and glob in addition to standard Anaconda distribution packages (Robitaille et al. 2013; Price-Whelan et al. 2018; Bradley et al. 2019; Anaconda Python Distribution 2021). Photutils version 0.7.2 was used for source extraction, including source deblending, aperture photometry, and error propagation from the background estimation process via the image segmentation package.

Astroquery was used to interface with astrometry.net (Lang et al. 2010) to plate solve each frame from the telescope so that astronomical coordinates would be available in the file. After source detection in pixel and sky plane coordinates, stellar sources were matched with the SDSS12 catalog (Alam et al. 2015). Astroquery was also used to query the asteroid ephemeris from the Minor Planet Center to confirm the target was properly identified in each frame. Traditional in-frame aperture photometry was performed for each frame in each filter, as outlined below. For a more in-depth review of in-frame photometry, differential photometry, and aperture photometry methods, please refer to Howell (2006).

A local background map was computed using a two-dimensional Photutils filter after masking point sources in the image. Aperture sizes for each image were selected using point spread function information measured by Photutils. The number of counts within each aperture was summed and the estimated local background subtracted. The resulting net number of counts was divided by the exposure time to provide a flux from each star and the target asteroid. Instrumental magnitude was then calculated for each catalog star and the median difference between the instrumental magnitude and catalog magnitude for all in-frame stars provides the zero point magnitude. Each frame included at least 50 stars to make this estimation more robust. The zero point magnitude for each frame was added to the instrumental magnitude estimated for the target asteroid in order to calculate the calibrated SDSS magnitude of the target used for constructing the lightcurves. Using this method of in-frame photometry makes the brightness measurements more robust to atmospheric variation throughout the night and allows night-to-night comparisons.

## 2.2. Lightcurves and Rotation Period

Figure 1 shows the four-color lightcurve of 1998 OR2 from 2020 April 16 and 17. For each filter and each night separately, the median magnitude was subtracted from the individual calibrated magnitude data points to produce a relative magnitude value. Lightcurves generated with relative magnitude values can be compared across photometric filters in most cases, allowing us to combine multiple filters in our period estimation.

Using the combined Sloan g', r', and i' relative magnitude lightcurves from the two nights, the rotational period of 1998 OR2 was estimated using the Lomb-Scargle periodogram (Lomb 1976; Scargle 1982) capabilities of Astropy (Robitaille et al. 2013; Price-Whelan et al. 2018). The Sloan z' filter data was not used for period estimation due to its low SNR compared to the other filters. The results of the Lomb-Scargle periodogram are shown in Figure 2 which shows the power present in each potential frequency of rotation (cycles per hour) ranging from one- to ten-hour periods. Two peaks of roughly equal power are found as potential rotational frequencies at 0.2855 and 0.2424 cycles per hour, corresponding to periods of 3.503 h and 4.126 h, respectively. The 3.503-hour period is a poor fit for 1998 OR2 since folding the asteroid's lightcurve on this period results in an ill-defined lightcurve. The 4.126-hour period has slightly more power in the peak than the alternative and is also closer to the rotation periods published in Koehn et al. (2014) and Warner & Stephens (2020). We use the half width at half maximum of the peak power to estimate the uncertainty in this measurement as is common for periodogram estimates (Bretthorst 2003; VanderPlas 2018). The final rotational period estimate for 1998 OR2 is 4.126 ± 0.179 hours.

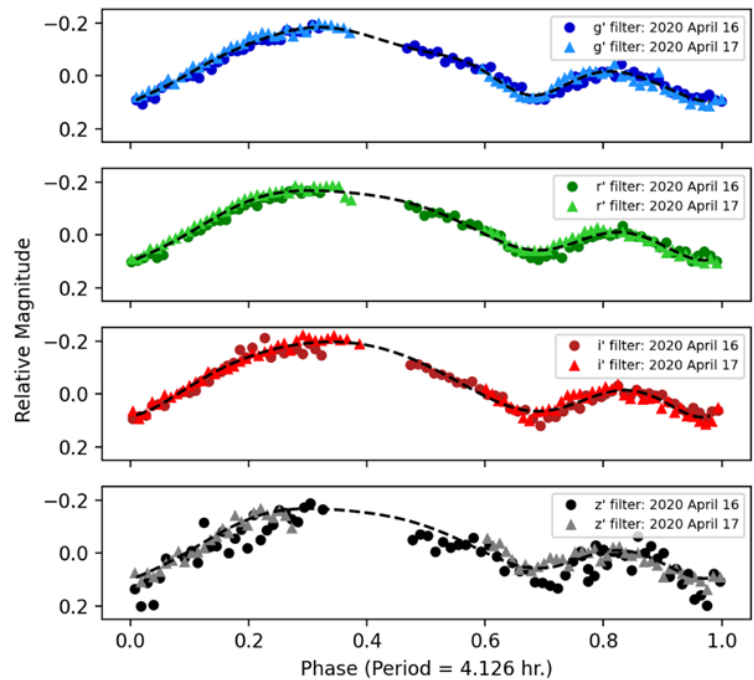

*Figure 1. Phase-folded lightcurves of near-Earth asteroid (52768) 1998 OR2 from LEO 0.5-meter telescope in four Sloan filters. Error bars are equivalent or smaller than the points used to display the data. The lightcurve and rotation period is consistent with those published by Koehn et al. (2014) and Warner & Stephens (2020).*

The Lomb-Scargle estimated rotational period of (52768) 1998 OR2 was confirmed by plotting the phased lightcurve in four filters across 2020 April 16 and 17 UTC in Figure 1. A period of 4.126 hours was used for phasing the lightcurves and the two brightness minima of the lightcurve were found to be of similar amplitude which matches the trend seen in Warner & Stephens (2020) on 2020 April 13 and 14. The two nights of data overlap, indicating that the period is well constrained.

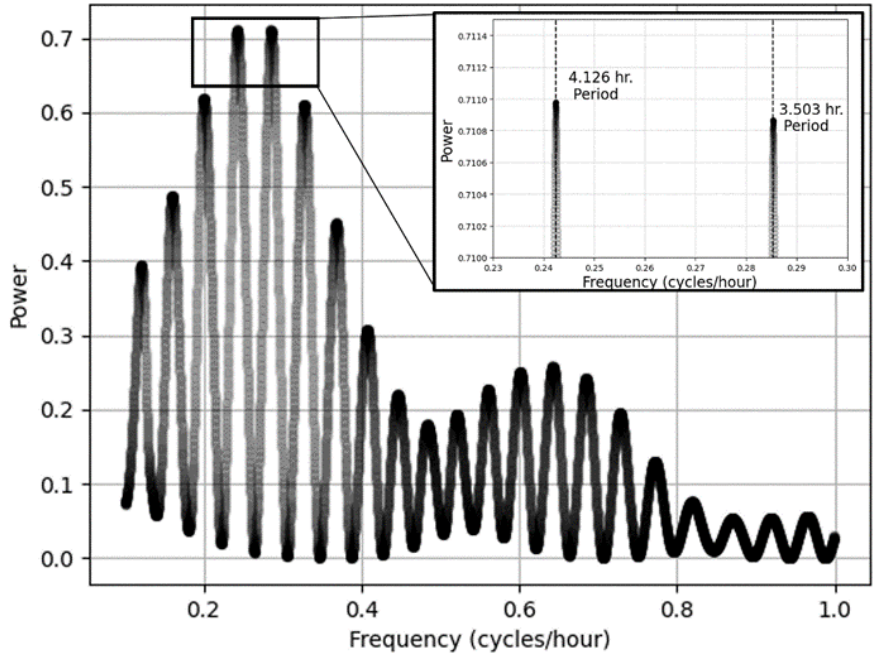

Figure 2. Lomb-Scargle periodogram of possible rotational periods for (52768) 1998 OR2. Two peaks of roughly equal power exist at 0.2855 and 0.2424 cycles per hour corresponding to periods of 3.503 and 4.126 hours, respectively.

3. SPECTROSCOPIC STUDY

   3.1. Visible Wavelength Observations and Data Reduction

Visible spectra (0.45 – 0.93 μm) of 1998 OR2 were collected using the Robotic Automated Pointing Telescope for Optical Reflectance Spectroscopy I (RAPTORS I) on 2020 April 16 UTC. The observational circumstances for these and the near infrared (NIR) spectra taken with NASA's Infrared Telescope Facility (IRTF) are shown in Table 2. The 0.61-m, f/4.64 RAPTORS I telescope is equipped with a 1,056 x 1,027 detector with 13 μm pixels. This results in a roughly 16 x 16 arcminute field of view with 0.9" px$^{-1}$ plate scale. A 30 line mm$^{-1}$ grating is mounted in a filter wheel in front of the CCD which gives a 14.6 nm px$^{-1}$ resolution and a spectral resolution of R~30 at 450 nm. Dark, bias, and flat-field frames in the diffraction grating were obtained on April 16 with RAPTORS I.

*Table 2. Observational circumstances for spectrometric observations*

| Telescope | Date UTC | Start UTC | End UTC | Airmass | V. Mag | α (deg) | Wavelength Range (μm) | No. of Img. |
|---|---|---|---|---|---|---|---|---|
| RAPTORS I | 2020 April 16 | 03:49 | 06:31 | 1.07 – 1.73 | 12.98 | 105.6 | 0.45 – 0.93 | 562 |
| NASA IRTF | 2020 Feb. 28 | 08:44 | 09:02 | 1.11 – 1.13 | 14.9 | 40.2 | 0.78–2.49 | 18 |

The spectral resolution was initially calculated using the grating equation and was refined with observations of the H-α line at 656 nm in Seyfert galaxy M106. The resolution was empirically confirmed by observing bright main belt asteroids and comparing to archival data in NASA's PDS Data Ferret.[2] The resulting slitless spectrometer system can observe targets as faint as 14th magnitude with a full night of observations.

Spectral data processing was performed using a reflectance spectroscopy pipeline written in Python which uses Source Extractor, the Smithsonian Astrophysical Observatory's DS9, pyds9, sewpy, and Skyfield in addition to standard Anaconda distribution packages (Bertin & Arnouts 1996; Joye & Mandel 2003; Montesano et al. 2016; MegaLUT 2017; Rhodes 2019; Anaconda Python Distribution 2021).

The pipeline allows the user to specify the spectral resolution, wavelength for normalization, target files, solar analog files, extraction box dimensions, background estimation box offset from the extraction box, the maximum allowed data value before CCD non-linearity, and the limit for the ratio of a point source's major- to minor-axes. More details on running the pipeline and what the code does at each step can be found in APPENDIX I: SPECTRAL PROCESSING PIPELINE.

The pipeline has two different modes for plotting the spectrum of the object. If the object is expected to have little spectral variation throughout the night – like we would expect for most asteroids – then the full night's worth of spectra are 3-sigma clipped to remove reflectance profiles with star contamination, windy conditions, or other poor observing conditions. The sigma clipped spectra are then median combined and the uncertainty is reported as the standard deviation of the sigma clipped spectra divided by the square root of the number of spectra used.

For objects with a large amount of spectral variation, such as small bodies with non-homogenous surfaces or near-Earth objects moving through a wide range of phase angles in one night, the median spectrum is calculated and the uncertainty is reported as the inner quartile range (IQR) of the night's reflectance at each wavelength. The IQR was chosen as a non-parametric way to understand the amount of spectral variation an object had throughout the time observed.

### 3.2. Near Infrared Observations and Data Reduction

The NIR spectrum (0.78–2.49 µm) of 1998 OR2 was acquired with the SpeX instrument on the NASA Infrared Telescope Facility (IRTF, see Rayner et al. 2003) and the processed data is taken from the public domain archive of data from the MIT-Hawaii Near-Earth Object Spectroscopic Survey (MITHNEOS). Observational circumstances for the MITHNEOS data are summarized in Table 2.

---

[2] https://sbntools.psi.edu/ferret/

The reduction of the MITHNEOS data is discussed in Binzel et al. (2019) and utilizes the "autospex" software tool to automate Image Reduction and Analysis Facility (IRAF) and Interactive Data Language (IDL) routines. The "autospex" tool runs on a single night of data at a time with a user checking the process at key points in the routine. The input to the software includes the raw asteroid and solar analog star images taken with AB nodding of the IRTF to allow sky subtraction.

The solar analog stars were observed as close to the asteroids as possible and at similar airmasses. Instead of standard stars observed at the same airmass as the asteroid for telluric and airmass corrections, the "autospex" tool for the MITHNEOS data uses the atmospheric transmission (ATRAN) model of Lord (1992) for final atmospheric corrections. The final reflectance spectrum is the asteroid flux divided by the solar analog flux corrected with the ATRAN model and is normalized to a wavelength of 1.215 μm for MITHNEOS data (Binzel et al. 2019). The combined visible-NIR spectrum of 1998 OR2 is shown in Figure 3.

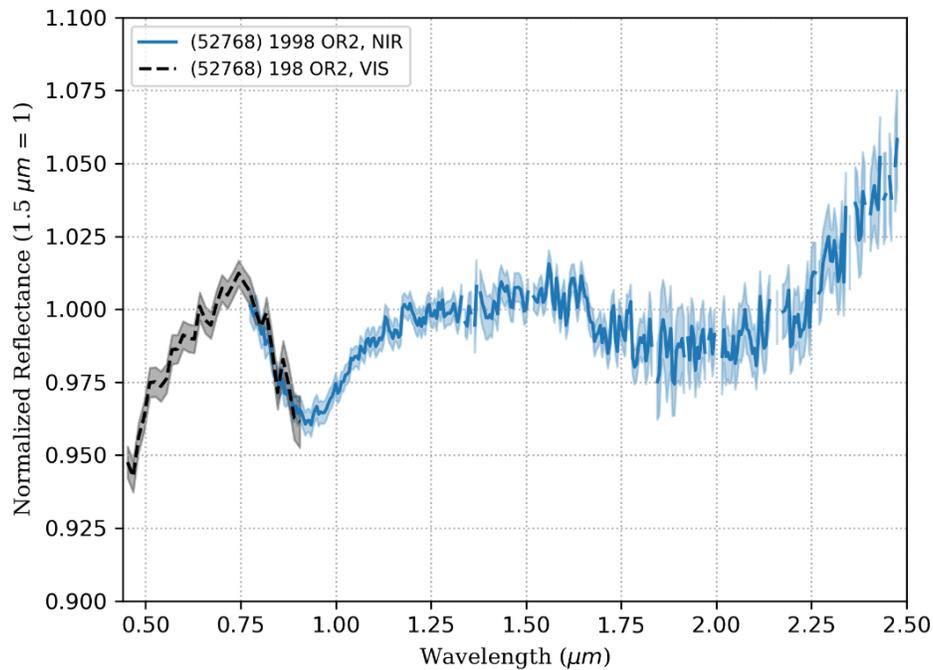

*Figure 3. Visible and near infrared spectra of near-Earth asteroid (52768) 1998 OR2. Visible data (0.45 – 0.93 μm) were taken with the RAPTORS I telescope from Tucson, Arizona on 2020 April 16 UTC. Near infrared data (0.78–2.49 μm) were obtained using the NASA IRTF on 2020 February 28 UTC and retrieved from the MITHNEOS survey archives. All data was normalized to 0.8 μm to provide overlap between visible and NIR data and then renormalized to 1.5 μm.*

### 3.3. Compositional Analysis

Figure 3 shows the VNIR spectrum of 1998 OR2 from RAPTORS I and NASA IRTF. The spectrum shows two absorption bands due to the presence of the minerals olivine and pyroxene: Band I at 0.926 ± 0.003 µm and Band II at 2.07 ± 0.02 µm. Band I has a depth of 4.5 ± 0.15% and Band II is 4.0 ± 0.21% deep with a Band Area Ratio of 1.13 ± 0.05. Diagnostic spectral band parameters measured from olivine and pyroxene absorption bands in the spectrum have been successfully used to constrain surface mineralogy, mineral chemistry and identify meteorite analogs (Reddy et al. 2015). For the combined visible and NIR spectrum of 1998 OR2, the spectral band parameters - band center, band depth, and band area - were measured using a Python code that utilizes only standard libraries such as those from the Anaconda Python Distribution (2021). The code follows the procedures described in Cloutis et al. (1986). The band center was calculated by dividing out the linear continuum and fitting a low-order polynomial over the bottom of the band. Band depth was measured as in Clark & Roush (1984) and is given as a percentage depth from the continuum to the reflectance of the band center.

The Band I parameters were measured twelve times using a third-order polynomial fit and sampling different ranges of points within the band. The Band II parameters were measured 13 times using the 2.45 μm upper limit as described in Sanchez et al. (2020), to account for the noisy data and missing points at the longest wavelength. The band area is calculated as the area between the continuum and the data curve. The wavelength range we used for these calculations was 0.45 – 2.45 μm. Previous studies (e.g., Cloutis et al. 1986; Gaffey et al. 1993; Dunn et al. 2010; Sanchez et al. 2020) have discussed that the BAR can be particularly sensitive to the wavelength range used. Sanchez et al. (2020) shows that BAR measurements for H chondrites can vary greatly if the incorrect wavelength range is used for the estimation. Although we use the best wavelength range [0.45 – 2.45 μm] for this measurement and follow methods laid out by previous studies, it is still possible that the BAR uncertainty could be larger than suggested by the variance in our repeated measurements. This could possibly explain 1998 OR2's location on the edge of the H chondrite region in Figure 4.

The reported values correspond to the average of these measurements for each parameter. Uncertainties for the band parameters are given by the standard deviation of the measurements. Table 3 lists the band center, band depth, band area ratio (BAR) values measured for 1998 OR2 including temperature corrected band measurements following the procedures of Sanchez et al. (2012). Due to 1998 OR2's close approach to the Earth, the asteroid's surface temperature is estimated to be relatively close to room temperature at 259 K, and temperature corrections have minimal impact on the band parameters. The Band I center is plotted against the BAR in Figure 4, showing the S-type asteroid subgroups from Gaffey et al. (1993).

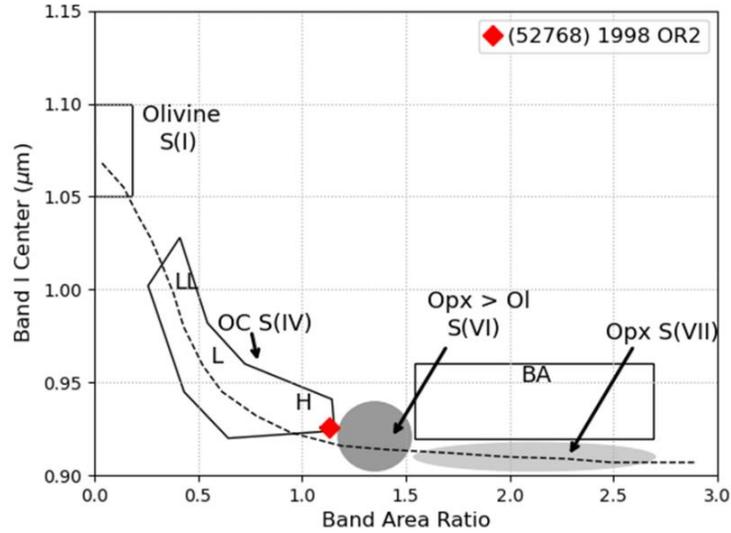

*Figure 4. Band I center vs. Band Area Ratio with select S-type asteroid subgroups from Gaffey et al. (1993). 1998 OR2 falls between the S(IV) OC "boot" and the S(VI) region. Error bars are on the order of the marker size.*

The position of 1998 OR2 on the Band I center versus BAR graph in the S(IV) region indicates the surface might be analogous to ordinary chondrites, more specifically H chondrites. While band depths are not diagnostic of surface mineralogy, it is interesting to note that 1998 OR2's absorption bands are subdued (4.5% and 4.0% for Bands I and II) compared to most S-complex asteroids (typically ~15 – 25% for Band I and ~5 – 10% for Band II; e.g., Sanchez et al. 2012). The spectrally derived olivine-pyroxene abundance ratio, ol/(ol+px), and the molar percent fayalite (Fa) and ferrosilite (Fs) were estimated using the appropriate equations from Dunn et al. (2010) and Sanchez et al. (2020). The summary of the band parameters and the estimated mineralogy is shown in Table 3.

*Table 3.* Observed mineralogy for (52768) 1998 OR2 derived from visible and NIR observations.

| Band I Center (μm) | Band I Depth (%) | Band II Center (μm) | Band II Depth (%) | Band Area Ratio (BAR) | Fayalite (mol%) | Ferrosilite (mol%) | ol/(ol+px) | $p_v$ Used | Temp. (K) |
|---|---|---|---|---|---|---|---|---|---|
| 0.926 [0.003] | 4.5 [0.15] | 2.07 [0.02] | 4.0 [0.21] | 1.13 [0.05] | 20.1 [2.3] | 18.2 [1.5] | 0.45 [0.04] | 0.164 | 259 |

*All values are reported after temperature corrections using a heliocentric distance of 1.18 AU during IRTF observations and assumed slope parameter, G=0.15. Uncertainties are shown in brackets below the measured value.*

To confirm the H chondrite affinity, we plot the estimated olivine and pyroxene chemistry of 1998 OR2 with the known ordinary chondrite types in Figure 5. The left diagram is adapted from Nakamura et al. (2011) and shows molar percent fayalite versus molar percent ferrosilite for several ordinary chondrites and the boundaries between the H, L, and LL chondrites. The right diagram is adapted from Dunn et al. (2010) and shows molar percent fayalite versus the olivine-pyroxene ratio. Both plots indicate a mineralogy consistent with H chondrites. The range of Fs/Fa uncertainties does not rule out L chondrites although H chondrites are better analogs considering Fa and the olivine/pyroxene ratio.

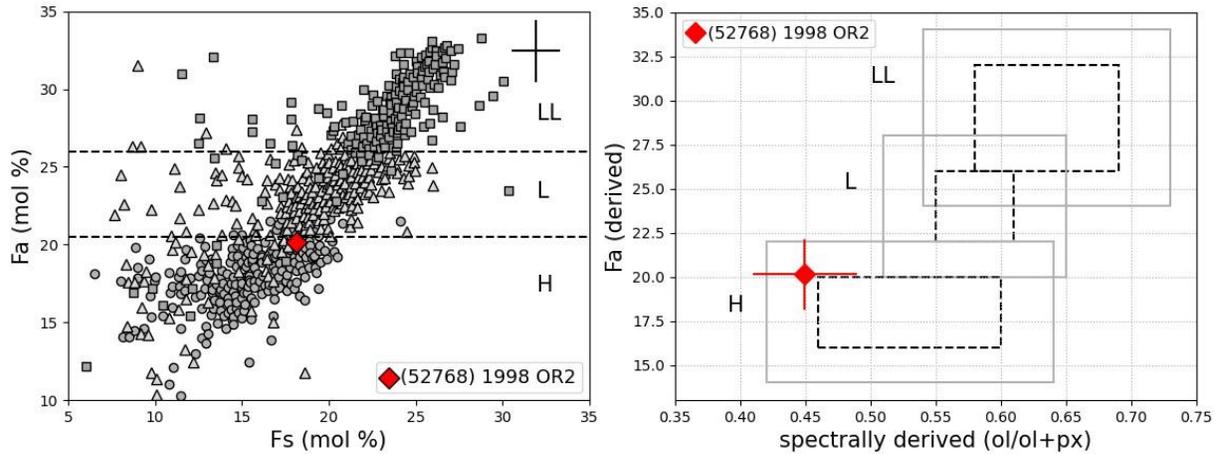

*Figure 5. (Left) Molar content of fayalite (Fa) vs. ferrosilite (Fs) derived for 1998 OR2 showing a mineralogy consistent with H chondrites. The error bars in the upper right corner correspond to the uncertainties derived by Sanchez et al. (2020), 2.0 mol% for Fa, and 1.4 mol% for Fs. Figure adapted from Nakamura et al. (2011). (Right) Molar content of fayalite (Fa) versus ol/(ol+px) ratio derived for 1998 OR2. Black dashed boxes represent the range of values for each ordinary chondrite subgroup and gray solid boxes represent the uncertainty in those measurements. Figure adapted from Dunn et al. (2010). Both plots indicate an H chondrite-like mineralogy.*

3.4. Taxonomic Classification

Thomas et al. (2014) found, using only the NIR spectrum from SMASS (Bus & Binzel 2002), that the closest taxonomic type to 1998 OR2 was the L-type from the Bus-DeMeo taxonomic system (DeMeo et al. 2009). L-type spectra are characterized by a steep slope in the visible wavelengths that flattens in the NIR. L-types typically have 1 μm and 2 μm features and can be uniquely distinguished from other taxonomic types by their positive PC1 values (DeMeo et al. 2009).

Others have found that 1998 OR2 is taxonomic type Xk (Binzel et al. 2019; Devogèle et al. 2020; Lazzarin et al. 2021). The Xk-type is characterized by a concave downward shape and a faint 0.8 to 1.0 μm feature. Xk-type spectra are red in the visible and are spectrally flat for visible wavelengths longward of 0.75 μm (DeMeo et al. 2009). The mean Xk spectrum from the MIT SMASS online tool, however, shows the Xk-type is red-slopped through the NIR wavelengths, making it less of a match for 1998 OR2 (Slivan 2013).

Classification of 1998 OR2 was done using the MIT SMASS online tool which also provides principal component values for the uploaded spectrum (Slivan 2013). The classification result required visual inspection to differentiate between Xk-, Xc-, Xe-, C-, Ch-, or Xn-type. The tool also returns residuals between the user-uploaded spectrum and the average spectrum for each type. For 1998 OR2, the Xk-type returned a residual of 0.188; Xc-type returned 0.125; Xe-type returned 0.150; C-type returned 0.062; Ch-type returned 0.046; and Xn-type returned a residual of 0.05. Although the Ch-type is the lowest residual, 1998 OR2 lacks the 0.7 μm phyllosilicate feature that is present in the Ch-types. The feature near 0.9 μm fits well for both Xk and Xn spectra, but the flat spectral slope, especially at wavelengths longer than ~1 μm, gives the Xn-type a much lower residual. We find that the Xn-type is the best taxonomic fit for 1998 OR2. The mean value and standard deviation for the two lowest residual taxonomies returned for 1998 OR2 from the MIT SMASS online tool (Xn- and Ch-type) are shown with the asteroid's spectrum in Figure 6 (right). Because 1998 OR2 is found to have an ordinary chondrite surface mineralogy, the S- and Q-type spectra are also shown for comparing 1998 OR2's suppressed mineral absorption bands to the more typical silicate asteroid spectra. This exercise illustrates the importance of using both visible and NIR data for taxonomic classification although the whole process could result in ambiguous results for weakly featured S-type asteroids as demonstrated here.

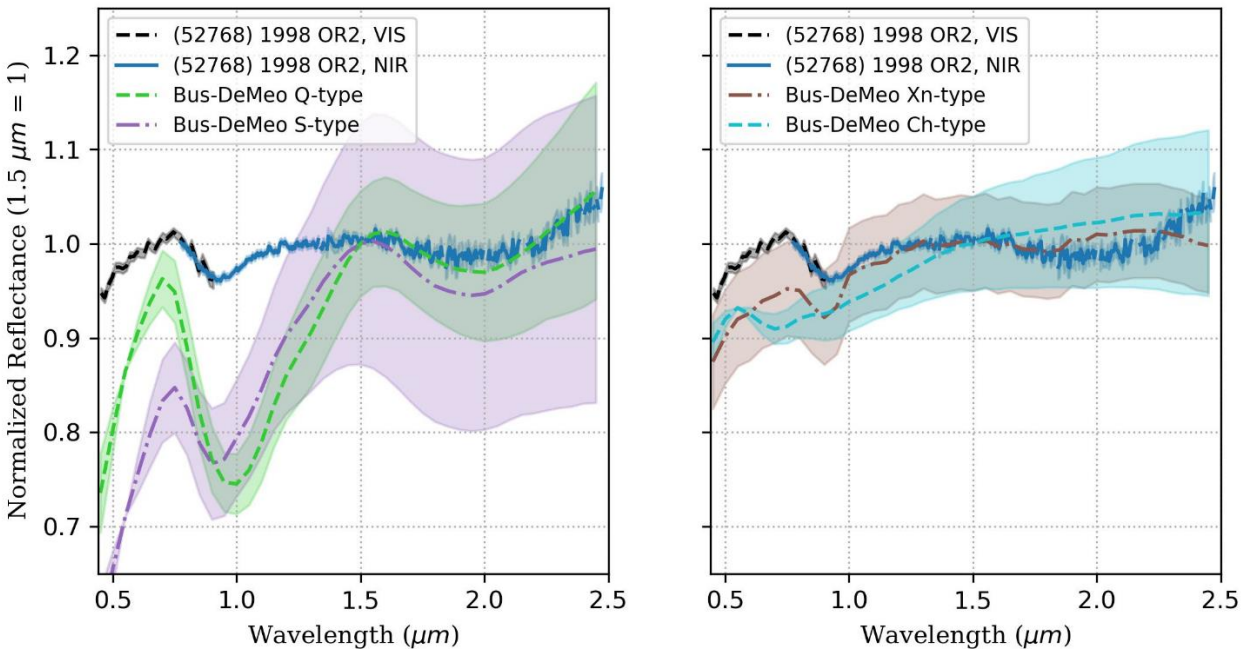

*Figure 6. Visible and near infrared spectra (0.45-2.49 μm) of near-Earth asteroid (52768) 1998 OR2 shown with (left) the average S- and Q-type spectra and (right) with the Xn- and Ch-type spectra from the Bus-DeMeo taxonomy. Shaded regions show the standard deviation of the average spectrum.*

The flat VNIR spectral slope alone is enough to indicate that 1998 OR2 does not have a metallic surface which would result in a reddened spectral slope (Ockert-Bell et al. 2010; Hardersen et al. 2011; Neeley et al. 2014; Cloutis et al. 2015; Cantillo et al. 2021; Sanchez et al. 2021). Radar measurements taken during the 2020 close approach also support 1998 OR2's potential S-type surface. These measurements include circular polarization ratio and radar albedo values which eliminate the possibility of a metal-rich surface and best match an S-type surface. Details surrounding these measurements and their implications are presented in APPENDIX II: METAL CONTENT. The disconnect between the mineralogy, which suggests an H chondrite meteorite analog, and the Xn taxonomic classification prompted us to explore the possibility of non-mineralogical factors (i.e., phase angle, grain size, space weathering or shock darkening/impact melt) affecting the spectrum of 1998 OR2.

### 3.5. Interpreting Spectral Characteristics

Spectroscopic observations of 1998 OR2 show suppressed band features and a flat spectral slope as well as an X-complex taxonomy. Many factors can affect the band parameters and slope for a reflectance spectrum including phase angle, grain size, space weathering, or the presence of shock darkened material. Here, we investigate each of these potential alteration processes in turn. The online MIT SMASS taxonomy classification tool provides measurements of principal component analysis (PCA) values for 1998 OR2 which were found to be PC1' = -0.388 and PC2' = -0.007 (Bus & Binzel 2002; Slivan 2013). These values for 1998 OR2 were plotted in Figure 7 along with the principal component values of 710 NEAs from Binzel et al. (2019). The α-line is shown, which divides traditionally silicate-rich S-type spectra from relatively featureless C/X-complex asteroid spectra. Objects below and left of the α-line typically do not show any prominent 1 μm or 2 μm feature (DeMeo et al. 2009). Space weathering and shock darkening trends from Binzel et al. (2019) and Reddy et al. (2014) are shown to illustrate potential alteration processes. The space weathering trend runs parallel to the α-line in PCA space.

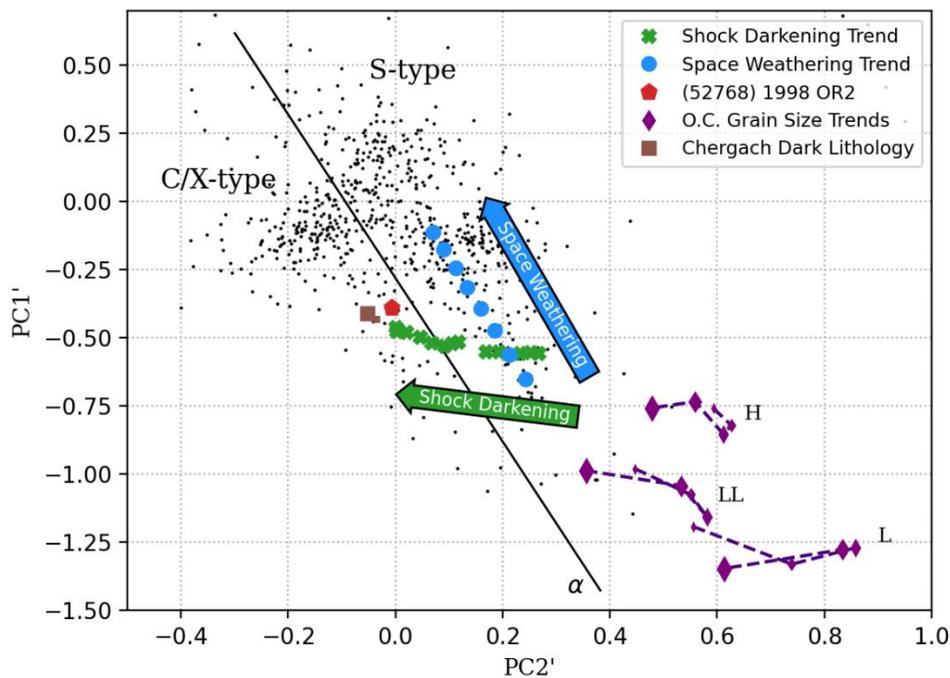

*Figure 7. Principal component analysis of 710 NEAs with the "α-line" boundary from DeMeo et al. (2009) which separates the S- and C/X-complexes. Space weathering and shock darkening trends are shown as adapted from Binzel et al. (2019) and Reddy et al. (2014). Ordinary Chondrite grain size trends are also shown for El Hammami (H), Viñales (L), and Chelyabinsk (LL) meteorites at 45-90 μm, 90-150 μm, 150-300 μm, 300-500 μm, and 500-1,000 μm grain sizes with increasing plot marker size. 1998 OR2 is shown at the end of the shock darkening trend, within the C/X-complex, despite having H-chondrite like mineralogy. Principal component values for the dark lithology of meteorite analog Chergach are shown for < 45 μm and 150 – 300 μm grain sizes.*

### 3.5.1. Phase Angle

Phase angle affects the slope and band depth of the observed asteroid spectrum and must be taken into consideration for NEAs which are frequently observed at large phase angles (Sanchez et al.

2012). 1998 OR2's spectrum is very flat, minimizing concerns about phase effects because the spectrum has not likely been phase reddened. Supporting this, the visible and NIR observations have good agreement in their wavelength overlap regions, despite a large phase angle difference between the visible and NIR observations. Band centers needed for mineralogical analysis are not largely affected by phase angle variations, especially at the moderate phase angle ($\alpha \sim 40°$) of the IRTF NIR observations where diagnostic band parameters were measured (Sanchez et al. 2012). The band centers and BAR, which are both minimally affected by phase angle, are the parameters used for calculating the surface mineralogy and mineral abundance as shown in Figure 4 and Figure 5.

Although band depth is not directly used in the calculations of the surface mineralogy, it is still linked to the surface properties of the asteroid. Band depth measurements of 12 S- and Q-type NEAs at different phase angles in Sanchez et al. (2012) show a range of Band I depths from roughly 15 – 25% across phase angles ranging from 2° - 87°. These measurements have corresponding changes in the Band I depth of less than 5% depth across all targets. These Band I depths are much larger than observed for 1998 OR2 and the phase angle induced depth changes still do not account for the low band depth seen in 1998 OR2.

Principal component values for the 12 NEAs were also plotted with boundaries for asteroid taxonomies in Figure 12 of Sanchez et al. (2012). This plot shows some of the targets changing between S-type subclasses, but none of the targets cross over the α-line into the C/X-type region as seen in Sanchez et al. (2012). This suggests that phase angle variations alone are not enough to produce the flat spectrum with suppressed band features or alter an S-type to a C/X-type taxonomic classification, like we see for 1998 OR2.

### 3.5.2. Grain Size

Grain size can affect the slope and band depths of asteroid NIR spectra (e.g., Cloutis et al. 2015). The true nature of asteroid regolith grain size is still highly debated and difficult to disentangle from other spectral effects. Some recent research points to smaller grain sizes (< 100 µm) being the most representative of asteroid surfaces with 3-40 µm being the most common (Nakamura et al. 2011; Reddy et al. 2015). Thermal inertia data from the WISE mission also supports a surface dominated by fine-grained regolith on moderate- to slow-rotators (P > 4 hrs.) such as 1998 OR2 (Hanuš et al. 2018).

Other studies have found that asteroids with diameters < 100 km tend to have millimeter to centimeter sized particles dominating the surface (Gundlach & Blum 2013). Another common regolith theory is that large particles may dominate the surface, but are themselves coated in fine (e.g. < 5 µm) dust grains (Craig et al. 2007; Vernazza et al. 2010). Although radar measurements from Bondarenko et al. (2020) indicate higher than average surface roughness at decimeter scales on 1998 OR2, this does not preclude the potential for the larger particles to be covered in fine-grained material dominating the spectrum. We also note that the radar measurements are sensitive to centimeter-scale roughness and will not be able to directly detect micron-sized particles on the asteroid surface.

To better investigate whether grain size could be an alternative explanation for 1998 OR2 falling on the C/X-complex side of the α-line, we use the principal component analysis of three different ordinary chondrites (O.C.s). Samples of El Hammami (H), Viñales (L), and Chelyabinsk (LL) meteorites were crushed and dry sieved to five different grain sizes: 45-90 μm, 90-150 μm, 150-300 μm, 300-500 μm, and 500-1,000 μm.

Spectra were obtained with an ASD LabSpec 4 Hi-Res spectrometer. This spectrometer has 3 nm resolution at 0.70 μm and 6 nm resolution at 1.40 and 2.10 μm. A 120 Watt quartz-tungsten bulb was used to illuminate the source at a 0° incidence angle while the reflected signal was measured at a reflected angle of 30°. Spectra were measured relative to a baseline Spectralon disk and corrected for dark current. Data were then processed with a python script that removed a known infrared feature of Spectralon and corrected for any detector offset at 1.0 and 1.8 μm (Kokaly et al. 2017; Cantillo et al. 2021).

The principal component (PC) values for each grain size were measured using the MIT SMASS online tool and were plotted in Figure 7. In all three meteorite types, the PC values between the four smallest grain sizes move mostly away from or parallel to the α-line in PC1' vs. PC2' space. Only the transition between the 300-500 μm and the 500-1,000 μm grain sizes moves the principal components directly toward the α-line in a way that could potentially change an S-type taxonomy to a C/X-type. None of the principal component value changes caused by different grain sizes altered the meteorite samples across the α-line, so we did not find that grain size altered the O.C. spectra in a way necessary to reproduce 1998 OR2's position in the PCA plot.

The principal component values for two grain sizes of Chergach, a shock darkened meteorite analog (see section 4) for 1998 OR2, were also measured in order to see the effects of grain size on 1998 OR2's position in principal component space. PC values for the dark lithology of this meteorite analog were measured for the two grain sizes of the meteorite analog and found to be PC1' = -0.4291 and PC2' = -0.0355 for the < 45 μm grain size; PC1' = -0.4091 and PC2' = -0.0520 for the 150 – 300 μm grain size. These values are shown in Figure 7 and are similar to 1998 OR2's values of PC1' = -0.388 and PC2' = -0.007. Different grain sizes of the meteorite analog of 1998 OR2 do not produce large changes in PC values, further ruling out the possibility that 1998 OR2's spectrum is a result of grain size alone. Given the above evidence, it is unlikely that grain size effects could produce the spectral characteristics of 1998 OR2.

### 3.5.3. Space Weathering

Lunar-style space weathering alters S-type asteroid reflectance spectra by lowing the albedo, suppressing band depths, and reddening the VNIR spectral slope (Hiroi et al. 2006; Gaffey 2010; Noguchi et al. 2011; Pieters & Noble 2016). The spectrum of 1998 OR2 is very flat even prior to continuum removal (as seen in Figure 3), suggesting that the surface has not been significantly altered by lunar-style space weathering. Principal component analysis trends of space weathering show that this alteration trend moves objects parallel to the α-line separating S-complex and C/X-complex asteroids (Binzel et al. 2019). Because the space weathering trend cannot alter asteroid principal component values across the α-line and 1998 OR2's PC values lie within the C/X-complex, it is unlikely that lunar-style space weathering alone would produce the spectrum of 1998 OR2.

### 3.5.4. Presence of Shock Darkened Material

The combined VNIR spectrum of 1998 OR2 was compared to the meteorites available in the Reflectance Experiment Lab (RELAB) using the Modeling for Asteroids (M4AST) online tool (Popescu et al. 2012; Birlan et al. 2016). M4AST uses a user-loaded, normalized reflectance spectrum and multiplies it by the median of a RELAB meteorite's absolute reflectance. Then, a chi-squared best-fit measurement is calculated for the match between the scaled asteroid spectrum and the RELAB meteorite spectrum. This process is repeated for all meteorites in the RELAB database and the top results are returned. For 1998 OR2, the top match was the McKinney L4 chondrite with shock darkened features (RELAB spectrum c2mh01, sample MH-CMP-001-P1; Milliken 2020). As noted in Figure 5 (left), the range of Fs/Fa uncertainties for 1998 OR2 does not rule out L chondrites, although H chondrites are better analogs considering Fa and olivine/pyroxene ratio. The spectrum of McKinney is shown in Figure 8 with the VNIR spectrum of 1998 OR2 scaled to the median albedo of the meteorite to match how results are returned from M4AST.

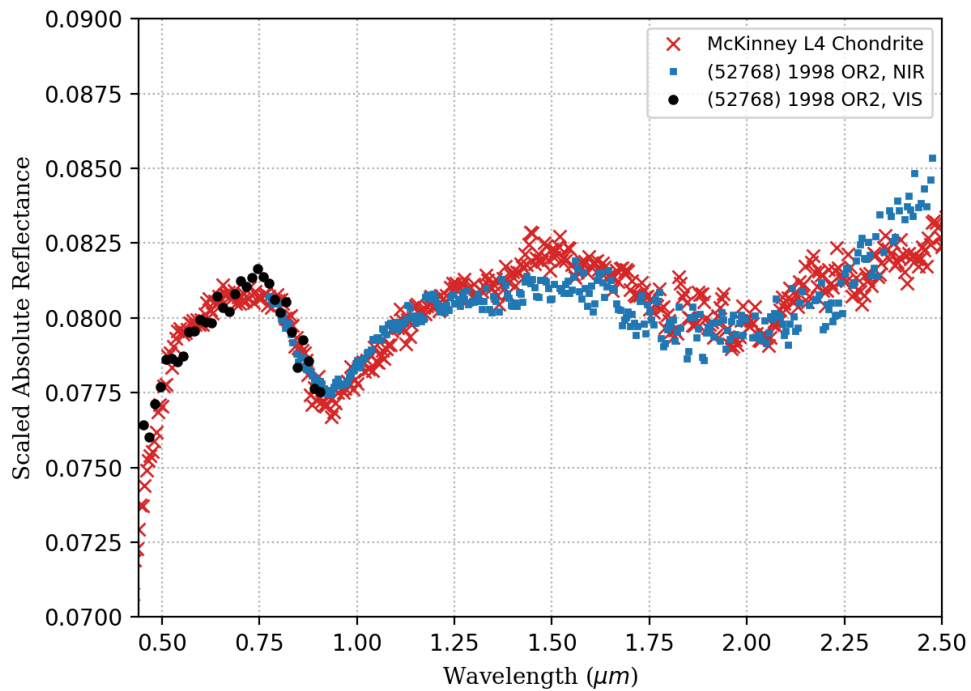

*Figure 8. Visible and near infrared spectra of (52768) 1998 OR2 shown with the RELAB McKinney shock darkened chondrite spectrum (RELAB c2mh01; Milliken 2020). The spectrum of 1998 OR2 was scaled to the median absolute albedo of the McKinney spectrum to match what was returned by M4AST.*

Because the above common alteration processes for asteroid spectra were eliminated, we interpret shock darkening as a viable explanation for the observed spectral characteristics of 1998 OR2. The presence of shock darkened material on the surface of the asteroid would result in a lowered albedo and suppressed mineralogical absorption features, but would have a minimal impact on the asteroid's spectral slope (Kohout et al. 2014; Reddy et al. 2014). The suppression of absorption bands moves the spectrum across the α-line in Figure 7 which traditionally separates spectra with a 2 μm absorption band from those without the 2 μm absorption band (DeMeo et al. 2009). By suppressing the absorption bands and moving spectra across the α-line, the presence of shock darkened material can alter spectra from S-complex into a C/X-complex taxon.

Previous works have discussed the possibility of 1998 OR2 having a heterogenous surface (Devogèle et al. 2020; Lazzarin et al. 2021). Polarization measurements correlating with a crater-like radar concavity on the asteroid's surface differ from the rest of the body implying either different scattering properties or a lower albedo in the surface feature (Devogèle et al. 2020). An earlier study of 1998 OR2 found the possibility of the asteroid having a negative G-parameter, which might indicate a low-albedo classification (Betzler & Novaes 2009). Further analysis of this crater-like concavity may provide more insight into the nature of this feature including the potential for it to be an impact crater which would support a hypothesis of shock darkening on 1998 OR2.

4. CONSTRAINING THE ABUNDANCE OF SHOCK DARKENED/IMPACT MELT MATERIAL

To further investigate the shock darkening hypothesis, and in order to constrain the amount of darkened material that could be present on the surface of 1998 OR2, the shock darkened H5 ordinary chondrite Chergach was prepared and analyzed in our lab as a potential meteorite analog. Chergach was identified as a potential meteorite analog for 1998 OR2 because of sample availability and the fact that it has been documented as an H5 ordinary chondrite with S3-S4 shock stage and minimal weathering (W0) (Weisberg et al. 2008). In addition, our mineralogical analysis shown in Figure 5 shows both H and L chondrites as possible analogs for 1998 OR2, although H chondrites are more likely if we consider both Fa and Fs chemistry and the olivine to pyroxene ratio. A section of the meteorite was loosely broken apart and the dark and light lithologies were manually separated. Dark lithologies were chosen that had only the darkest material and, as much as possible, contamination from light lithology and rust were removed. Due to limited sample availability, light lithology included both white and light-gray lithologies after all dark lithologies were removed.

Once good isolation of the desired lithology was achieved, the sample was crushed and dry sieved to a grain size of < 45 microns. This grain size was chosen because there is no clear answer as to what grain size best represents asteroid surfaces, and a full analysis of various grain sizes is not practical for this study due to the limited amount of meteorite material. The samples were analyzed using the same ASD spectrometer and methodology as the grain size study of the ordinary chondrite samples (Section 3.5.2). Laboratory spectra of Chergach were taken at a phase angle of 30° in order to closely match the observing conditions from the IRTF NIR observations (40.2°).

Mathematical aerial mixing was performed using the dark and light lithology spectra as endmembers to compare with the observed spectrum of 1998 OR2. All three spectra were continuum corrected by fitting a 3$^{rd}$ order polynomial to the peaks on either side of the Band I and

II absorption features at roughly 0.75, 1.5, and 2.5 μm. After the continuum was fit, the reflectance spectrum was divided by the continuum polynomial and the resulting spectrum was renormalized at 1.5 μm. The continuum was removed from 1998 OR2 and meteorite analog spectra with the same methodology to isolate band parameters and ensure that slope differences would not affect the compositional analysis. A more in-depth explanation of this continuum correction process can be found in Sanchez et al. (2020).

The continuum-corrected light and dark endmembers were iteratively scaled by different mixing ratios and added together before renormalizing again. A chi-squared analysis was used to determine the goodness of fit for each aerial mixture compared to the observed visible-NIR spectrum of 1998 OR2. The spectrum with the minimum chi-squared value was taken as the best fit for the aerial mixture.

Figure 9 shows the gradient of aerial mixing between Chergach endmembers with the darkest points corresponding to pure dark lithology and lightest points corresponding to pure light lithology, in 10% increments. The best-fit match with 1998 OR2 was determined using the minimum of a chi-squared calculation with one-percent increments and corresponds to 63% dark lithology and 37% light lithology.

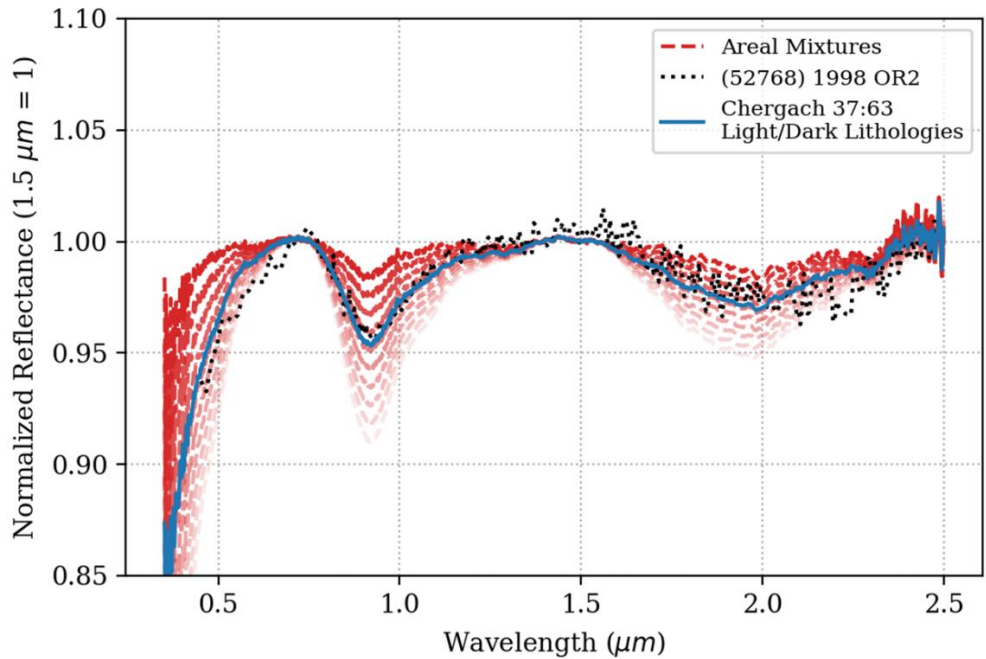

*Figure 9. Telescopic visible and near infrared spectrum of 1998 OR2 with laboratory spectra of Chergach, a shock darkened H5 ordinary chondrite. All Chergach and 1998 OR2 spectra were continuum removed prior to mixing and comparisons. Chergach light and dark lithology spectra were obtained as endmembers and were aerially mixed in different ratios to find the best match for 1998 OR2. The plotted gradient goes from pure dark lithology (darkest points) to pure light lithology (lightest points) in 10% increments. The best match for 1998 OR2 was found to be 63% dark lithology, 37% light lithology and is represented by the blue line.*

## 5. DISCUSSION

Our analysis of 1998 OR2 has shown that it has an S-type surface that has been altered by shock darkening or impact melt processes. This is supported by radar circular polarization measurements being close to S-type values (see APPENDIX II: METAL CONTENT), surface mineralogy matching that of H/L chondrites, PCA of 1998 OR2 being on the C/X-complex side of the α-line at the end of the shock darkening trend, and the asteroid's spectrum closely matching that of shock darkened H chondrite, Chergach, when mixed in a 37:63 light/dark lithology ratio.

The result that 1998 OR2 might be spectrally-dominated by shock darkened or impact melt material has direct implications for understanding the broader NEA population. For NEAs between 200 m and 5 km, S-type asteroids make up 50-60% of the population with closer to 70-80% representation in NEAs with the largest diameters, i.e., > 5 km (Binzel et al. 2019). Some recent studies, however, suggest that more primitive bodies (i.e., C- and D-types) may be more abundant in the smaller NEA population (i.e., < 200 m), but are less observed due to observational biases (Perna et al. 2018; Ieva et al. 2020). Ordinary chondrites, which make up ~85% of the meteorite collection (Grossman 2021), have been confirmed to originate from S-complex asteroids (Nakamura et al. 2011). As such, we might expect the percent of S-complex NEAs to be close to 85% in the size range of 1998 OR2 (200 m – 5 km), matching the ordinary chondrite representation seen in the meteorite collection. Other factors such as carbonaceous chondrite friability and atmospheric screening effects (Melosh 1997) favor ordinary chondrites reaching the surface, making the 85% value an upper limit on the percent of S-complex NEAs we should expect to see. One possible cause for the potential underrepresentation of S-complex NEAs, however, is that shock darkening processes are causing the S-complex taxonomy to be misidentified as weakly-featured taxonomies in the 200 m – 5 km size ranges.

This concept was originally investigated decades ago (see Britt & Pieters 1989, 1994, 1991), but has been gaining more attention since the 2013 Chelyabinsk event (Kohout et al. 2014; Reddy et al. 2014; Kohout et al. 2020a, 2020b). We provide here some of the first evidence of possible ordinary chondrite parent bodies hiding amongst the weakly-featured NEA population. Further research into shock darkening and impact melt trends on asteroids and meteorites is recommended to better rectify the percent of ordinary chondrites in the meteorite collection with the percent of S-type asteroids in the NEA population.

## 6. SUMMARY

We carried out detailed photometric and spectroscopic characterization of near-Earth asteroid (52768) 1998 OR2. Lightcurves were used to determine the rotation period of the asteroid and spectral band parameters were measured for mineralogical analysis. In-lab measurements of the shock darkened H chondrite meteorite analog, Chergach, were performed and an aerial mixing model was used to estimate the ratio of light and dark lithologies required to match 1998 OR2's spectrum. Below we summarize our findings and conclusions from these measurements and their implications for the NEA population and ordinary chondrite sources.

- Phased lightcurves across two nights of photometric observations of 1998 OR2 allow us to confirm the asteroid's 4.126 ± 0.179-hour rotation period.

- Combined visible and NIR spectroscopy allow us to measure the band parameters for 1998 OR2 and estimate its surface mineralogy. The asteroid's surface is estimated to have 20.1 ± 2.3% molar percent fayalite, 18.2 ± 1.5% molar percent ferrosilite, and an olivine-pyroxene ratio, ol/(ol+px), of 0.45 ± 0.04. These values agree with an H chondrite-like surface suggesting affinity to S/Q-complex asteroids.

- Absorption band depth of 1998 OR2 are subdued (4.5 ± 0.15% for Band I and 4.0 ± 0.21% for Band II) compared to typical S/Q types (~15 – 25% for Band I and 5 – 10% for Band II) suggesting a non-mineralogical process acting on the surface.

- Principal component values of 1998 OR2's VNIR spectrum fall on the C/X-complex side of the α-line. Taxonomic classification shows 1998 OR2 may be similar to the relatively rare Xn-type suggesting the limitation of taxonomy when classifying weakly featured S/Q-type asteroids.

- Radar albedo and circular polarization ratios were found to best agree with S-type asteroids (see APPENDIX II: METAL CONTENT). These measurements are able to rule out a metallic surface (M-type) and make a C- or X-type surface improbable.

- Aerial mixing modeling of shock darkened H5 chondrite Chergach showed good agreement with 1998 OR2 when mixed in a 37:63 light to dark lithology ratio.

- Mineralogical, albedo, and radar surface properties all support that 1998 OR2 may be an S-type asteroid that has been shock darkened, reducing its geometric albedo, suppressing mineralogic bands, and altering the taxonomic type from S-complex to the C/X-complex.

- This study presents the first possible evidence of shock darkening on the surface of a near-Earth asteroid. If shock darkening is relatively common among the NEA population, this could account for the mismatch between the percent of S-type asteroids in the NEA population and the percent of ordinary chondrites in the meteorite collection.

Acknowledgments

This research work was supported by NASA Near-Earth Object Observations grant NNX17AJ19G (PI: Reddy). This research has made use of the Small Bodies Data Ferret (http://sbn.psi.edu/ferret/), supported by the NASA Planetary System. Part of the data utilized in this publication were obtained and made available by the MITHNEOS MIT-Hawaii Near-Earth Object Spectroscopic Survey. Observations reported here were obtained at the Infrared Telescope Facility, which is operated by the University of Hawaii under Cooperative Agreement NCC 5-538 with the National Aeronautics and Space Administration, Science Mission Directorate, Planetary Astronomy Program. Taxonomic type results presented in this work were determined, in whole or in part, using a Bus-DeMeo Taxonomy Classification Web tool by Stephen M. Slivan, developed at MIT with the support of National Science Foundation Grant 0506716 and NASA Grant NAG5-12355.

The authors would like to acknowledge the significant cultural role Maunakea has with the Native Hawaiian community and how much we benefit from telescopes occupying their most sacred site.

We recognize the urgent suggestions of papers such as Kahanamoku et al. (2020) and Prescod-Weinstein et al. (2020) for observational astronomers to familiarize themselves with Hawaiian history and culture and recommend "Understanding Mauna Kea: A Primer on Cultural and Environmental Impacts" (2020) from the University of Hawai'i at Mānoa's William S. Richardson School of Law as a potential starting point.

Much of our data acquisition and processing was done in Tucson, Az. which is on the land and territories of Indigenous Peoples. We acknowledge our presence on the ancestral lands of the Tohono O'odham Nation and the Pascua Yaqui Tribe who have stewarded this area since time immemorial.

## 7. APPENDIX I: SPECTRAL PROCESSING PIPELINE

When run, the first solar analog image is displayed for the user to click near the zeroth order of the solar analog star. Source Extractor is used to determine all point sources in the image with an ellipse axis ratio smaller (more circular) than the user-specified limit and the closest point source to the user's click is used as the solar analog. For each solar analog image, the extraction box and background box are drawn on the image in a DS9 window so that the user can monitor which object is being measured. The program monitors the position of the extraction box and will request that the user select the target again if the object moves more than 50 pixels from the original position or no point source is detected within a 50 x 50 pixel box around the clicked position.

For each image, the "zero" wavelength position is determined by the centroid and wavelength is calculated as the number of pixels from the centroid times the user-given wavelength resolution. For each pixel in the width of the extraction box, the counts in a column that is the height of the extraction box are summed together and recorded for that position's calculated wavelength. The background is estimated using the median of all pixels inside the background-estimation box to be more robust against hot pixels and star contamination. This median background value is multiplied by the height of the extraction box and subtracted from each column's sum. Finally, each spectrum is normalized to the user-defined wavelength. This procedure is done for all solar analog images and the results are median combined and recorded.

The extraction process is repeated for each individual target image with a second initial click on the first target image from the user. For each target image, the resulting normalized spectrum is divided by the median-combined solar analog spectrum to produce a reflectance spectrum. Individual target spectra also have the brightness of the zeroth order point source output by Source Extractor to make uncalibrated lightcurves. The brightness is the sum of the counts within the Source Extractor aperture with the Source Extractor background estimate subtracted and divided by the exposure time to give a flux measurement (Bertin & Arnouts 1996).

The median-combined solar analog and each individual object image's reflectance and raw sum of counts per wavelength are recorded in a workbook. Meta-data such as the time since the first image, the date, centroid pixel position, and object name are also recorded in this workbook.

A second Python script is used for combining individual spectra, converting measured values, and making plots out of the raw data. The flux value measured with Source Extractor from the zeroth-order point source is used to calculate a nominal Sloan g' filter magnitude using equation 2.

$$m_{nominal} = -2.5 * \log_{10}(flux) + zero\ point \quad (2)$$

Here, the zero point value is a one-time estimated offset measured for the RAPTORS I system in order to move the instrumental magnitude close to the true brightness of a target. This one-time calibration was performed by simultaneously observing a target with the LEO telescope and RAPTORS I. The zero point value was then adjusted such that the nominal magnitude output for RAPTORS I matched the LEO telescope's in-frame photometry measured magnitude. This does not indicate that the nominal magnitude for RAPTORS I is calibrated, but this is done to make the results more understandable. Lightcurves are generated using the nominal Sloan g' magnitude and the time since the first image. These lightcurves are most useful for understanding where in the asteroid's rotation phase the spectra were obtained and are not presented here since higher quality, calibrated lightcurves are available from the LEO telescope.

## 8. APPENDIX II: METAL CONTENT

The flat spectral slope of 1998 OR2's spectrum suggests the asteroid is not metal rich since metallic asteroids typically have highly reddened slopes (Ockert-Bell et al. 2010; Hardersen et al. 2011; Neeley et al. 2014; Cloutis et al. 2015; Cantillo et al. 2021; Sanchez et al. 2021). Using radar measurements, however, we can further confirm 1998 OR2's composition is non-metallic. The opposite circular polarization (OC) radar albedo of 1998 OR2 was measured at 0.11 and the polarization ratio was estimated at 0.39 (Bondarenko et al. 2020).

These radar measurements together best match the S-type taxonomy which is the only member of the Tholen taxonomy that reliably has polarization ratios above 0.3 in Magri et al. (2007). The C- and M-type asteroids in the same study have an average polarization ratio of 0.098 and 0.153 respectively, whereas S-types have an average of 0.198 with a potential bimodality showing a second peak close to 0.3, better matching the estimated polarization ratio of 0.39 for 1998 OR2. More recent studies show somewhat ambiguous identification between X-, C-, and S-type taxonomies due to the wide range of polarization values measured, but with S- or Q-type still being the closest match to 1998 OR2 with average polarization ratios of $0.250 \pm 0.151$ and $0.274 \pm 0.107$ respectively (Benner et al. 2008; Virkki et al. 2014; Virkki & Muinonen 2016).

The average radar albedos of C- and S-type asteroids are $0.135 \pm 0.056$ and $0.142 \pm 0.046$, respectively, which are similar to one another. M-type, metallic asteroids, and X-type asteroids however have high radar albedos, averaging at $0.294 \pm 0.135$ and $0.220 \pm 0.194$, much higher than observed for 1998 OR2 at 0.11. These radar measurements make it improbable that 1998 OR2 is a metallic asteroid or that the subdued absorption bands and taxonomic classification are likely the result of a high metal content on the surface of the asteroid. Together, the available measurements imply that 1998 OR2 might have an S-type surface at radar wavelengths.